\documentclass[prb,reprint,showpacs,preprintnumbers,amsmath,amssymb,floatfix]{revtex4-1}
\usepackage{graphicx}
\usepackage{color}

\begin{document}

\title{Interplay of bulk and interface effects in the electric-field driven transition in magnetite}

\author{A.~A.~Fursina$^{1}$, R.~G.~S.~Sofin$^{2}$, I.~V.~Shvets$^{2}$, D.~Natelson$^{3, 4}$}

\affiliation{$^{1}$ Department of Chemistry, Rice University, 6100 Main St., Houston, TX 77005}
\affiliation{$^{2}$ CRANN, School of Physics, Trinity College, Dublin 2, Ireland}
\affiliation{$^{3}$ Department of Physics and Astronomy, Rice University, 6100 Main St., Houston, TX 77005}
\affiliation{$^{4}$ Department of Electrical and Computer Engineering, Rice University, 6100 Main St,.Houston, TX 77005}

\date{\today}


\begin{abstract} 
Contact effects in devices incorporating strongly-correlated electronic materials are comparatively unexplored.  We have investigated the electrically-driven phase transition in magnetite (100) thin films by four-terminal methods. In the lateral configuration, the channel length is less than 2~$\mu$m, and voltage-probe wires $\sim$100~nm in width are directly patterned within the channel.  Multilead measurements quantitatively separate the contributions of each electrode interface and the magnetite channel. We demonstrate that on the onset of the transition  contact resistances at both source and drain electrodes and the resistance of magnetite channel decrease abruptly. Temperature dependent electrical measurements below the Verwey temperature indicate thermally activated transport over the charge gap. The behavior of the magnetite system at a transition point is consistent with a theoretically predicted transition mechanism of charge gap closure by electric field.

\end{abstract}

\maketitle


The complex iron oxide, magnetite, Fe$_3$O$_4$, is an example of strongly correlated 3d-electron systems \cite{1998_Tokura_MIT_review}. It has been known for decades that bulk magnetite undergoes a first-order metal-insulator transition (two-order-of-magnitude change in electrical resistivity) at the so-called Verwey temperature, $T_{\mathrm{V}}\sim$120 K, accompanied by a structural transformation \cite{1939_Verwey_first_Nature}. 

Efforts on magnetite characterization are numerous in the seventy years since the discovery of the Verwey transition, including thorough investigations of its electrical properties \cite{2002_Walz_review, 1954_mag_el_lowT, 2007_planar_Hall, 1982_el_trans_MO, 2007_spin_PES_review} supported by theoretical calculations of electronic structure \cite{1987_polaron_cond,2004_CO_OO,1984_highT_band_struc_calc}.  Recent advances in nanofabrication and film growth allow electrical characterization at previously inaccessible scales, leading to the recent discovery of an electric field driven transition:  Magnetite films or nanoparticles below $T_{\mathrm{V}}$ experience a transition from an insulating state to a state with much lower resistance upon application of a sufficiently high voltage \cite{Our_magnetite_2008, 2008_APL_HAR_Cr, 2009_PRB_hyster}. The switching voltage scales linearly with the channel length suggesting an electric-field driven transition.

The key point of these experiments was an examination of magnetite films or nanoparticles between two electrodes separated by only several hundreds of nanometers or less. In this configuration the electric field needed to drive the transition was accessible at relatively low voltages, thus preventing both excessive heating and damaging of the sample.  We proved the observed switching not to be an artifact of heating \cite{Our_magnetite_2008, 2009_PRB_hyster}, in contrast to previously observed transitions in magnetite driven by Joule heating of the samples above $T_{\mathrm{V}}$ under bias \cite{ 1969_Tinduced_MIT_Fe3O4_1, 1969_Tinduced_MIT_Fe3O4_2}. 

The downside of such small channel length experiments is an unavoidable, dominant contribution of the contacts, which prevents direct insight into the properties of magnetite before and after transition.  By fitting our data for two-terminal devices with different channel lengths it was demonstrated that contact resistance of Au/magnetite interfaces comprises more than 70 \% of the total resistance \cite{Our_magnetite_2008}.  Upon testing several different contact metals (Au, Pt, Cu, Fe and Al), copper showed the lowest contact resistance with magnetite film \cite{2009_PRB_hyster}.  Even with a Cu contacting layer, however, the contribution of the contacts to the total two-terminal device resistance cannot be neglected.

One of the most effective ways to differentiate between bulk and interface effect is to make multilead measurements.  To date no such experiments have been performed to study the recently discovered electrically driven transition in magnetite.
In this paper we perform four-terminal experiments in a lateral electrode configuration using magnetite thin films.  The channel length is less than 2 $\mu$m and voltage-probe wires $\sim$100 nm in width are directly inserted into the channel. These multilead experiments quantitatively and unambigously separate the role of each interface and the magnetite channel. For the first time we study the changes in contact and channel resistance contributions at the onset of electric-field-driven transition in magnetite. Results indicate that at the transition point {\it both} contact resistances and the resistance of the magnetite channel decrease abruptly.  By doing temperature-dependent electrical measurements below $T_{\mathrm{V}}$ we trace the thermally activated transport over the charge gap in magnetite and provide an insight into the  transition mechanism  in this system.


The Fe$_{3}$O$_{4}$ (100) thin films (thickness: 50-100nm) used in the present study were grown on (100) oriented MgO single crystal substrates  using oxygen plasma assisted molecular beam epitaxy system (DCA MBE M600) with a base pressure 2$\times$10$^{-10}$ Torr. The substrates were cleaned in-situ at 873~K in 5$\times$10$^{-6}$ Torr oxygen for two hours. 

Reflection high energy electron diffraction, RHEED, (STAIB Instruments) was used to monitor the growth mode and growth rate (0.3 $\mathrm{\AA}$/s). Room temperature Raman spectroscopy (performed in the backscattering configuration using Rainshaw 1000 Micro Raman system), High resolution X-Ray diffraction measurements using a multi-crystal high-resolution X-ray diffractometer (HRXRD, Bede-D1, Bede, UK), and low temperature four probe resistance measurements were performed to establish the crystal structure and stoichiometry of the Fe$_{3}$O$_{4}$ phase \cite{Shvets_backscat,Shvets_high_res_Xray}.  

\begin{figure}
\begin{center}
\includegraphics[clip,width=8cm]{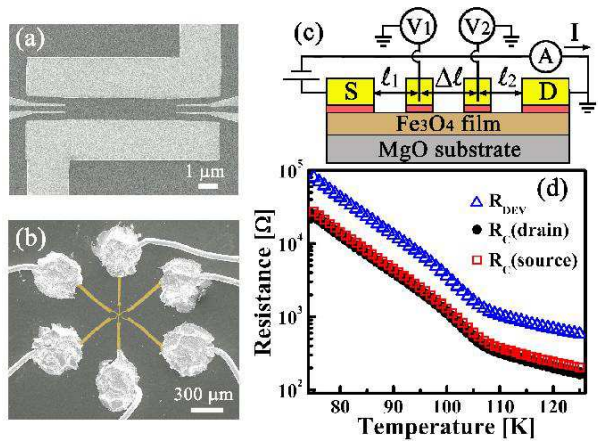}
\end{center}
\vspace{-5mm}
\caption{\small (color online) (a) SEM image of the device for four-probe measurements showing source and drain leads and two pairs of voltage probes within the channel. (b) Colored SEM image demonstrating electrical contacts to $\mu$m-size Au pads with further In soldering to attach Au wires. (c) Schematics of electrical circuit of four-probe measurements. Letters S and D  denote source and drain contact, respectively. Contacts are made of 6~nm Cu adhesion layer (reddish) and  10-20~nm cover layer of Au (yellow). (d) Temperature dependence of the low-bias resistance of magnetite channel ($R_{\mathrm{DEV}}$) and corresponding contact resistances ($R_{\mathrm{C}}$) at source and drain electrodes.
}
\label{fig1}
\vspace{-3mm}
\end{figure}

Devices for two- and four-terminal measurements were prepared by electron beam lithography (Jeol-6500 SEM).  A channel length of 400~nm - 1.9~$\mu$m is defined by two 10~$\mu$m wide source and drain leads. One or two pairs of voltage-probe leads were directly patterned within the channel (Fig.~\ref{fig1}a).  The contacts  were fabricated by the electron-beam deposition. In a typical experiment 6~nm of Cu (the best adhesion to magnetite film and lowest contact resitance out of studied contact metals \cite{2009_PRB_hyster}) and 10-20~nm  cover layer of Au were deposited. The leads were connected to micrometer-size pads (300 $\times$ 300 $\mu$m) to which Au wires are attached by In soldering (Fig.~\ref{fig1}b) and then connected to  external contacts of the puck. The puck was placed into the chamber of a Quantum Design Physical Property Measurement System (PPMS model 6000)  for variable temperature measurements (300~K - 80~K).  The lower temperature bound of these measurements is limited by the increasing switching voltages and concerns about device damage as $T$ is further decreased.


Electrical characterization of the devices was performed by standard four-terminal methods using a semiconductor parameter analyzer (HP 4155A). The schematic of device electrical connections  is presented in Fig.~\ref{fig1}c. The voltage, $V_{\mathrm{out}}$, is applied to the source lead with the drain grounded, and current flowing through the channel is recorded.  The pair of voltage probes, directly inserted into the channel between source and drain leads, senses voltages $V_1$ and $V_2$.  A voltage drop in the channel without the contact contribution is then calculated as $\Delta V=V_{1}-V_{2}$.  Only one pair of voltage probes (either left or right, see Fig.~\ref{fig1}a) is active in a given measurement, with the second pair being intact.  Having two pairs of voltage probes allows two independent sets of measurements (one for each pair of voltage probes) in a given channel, to verify data consistency in these four-terminal devices.  As was demonstrated in our previous paper, sweeping voltage in a continuous staircase mode leads to the overheating of the sample and appearance of a hysteresis in forward and reverse bias sweeps \cite{2009_PRB_hyster}. To minimize Joule heating of the channel, the voltage was always swept in a {\it pulsed} regime with the shortest available pulse duration (500 $\mu$s) and with pulse period $>$ 5 ms.  This pulse measurement procedure greatly reduced apparent hysteresis in the transition as a function of bias sweep \cite{2009_PRB_hyster}.

At any $V_{\mathrm{out}}$, voltage first drops at source electrode/Fe$_3$O$_4$ interface ($V_{\mathrm{C}}$(source)).  We assume that at low source-drain biases the contact interface contributions are dominated by an Ohmic contribution, $R_{\mathrm{C}}$(source)$\equiv V_{\mathrm{C}}$(source)/$I$.  Then, in the assumption of a homogeneous film (medium) between the  electrodes, voltage {\it linearly} drops across the channel, and two values are recorded at the two locations of the voltage probes.  The remaining potential drop to zero volts (grounded drain electrode) occurs at Fe$_3$O$_4$/drain electrode interface ($V_{\mathrm{C}}$(drain)$\rightarrow$ $R_{\mathrm{C}}$(drain)).  Conventionally, the total device may be represented as a voltage drop over three resistors in series, $R_{\mathrm{C}}$(source), $R_{\mathrm{DEV}}$ and $R_{\mathrm{C}}$(drain). By knowing the geometrical characteristics of our devices from SEM images (i.e., $\ell_1$, $\Delta \ell$, and $\ell_2$, see Fig.~\ref{fig1}c) we can calculate the values of all three voltage drops and, by dividing over measured current, corresponding resistances.

An example of the temperature dependence of  $R_{\mathrm{DEV}}$, $R_{\mathrm{C}}$(source) and $R_{\mathrm{C}}$(drain), calculated this way  at $V_{\mathrm{out}}$ = 100mV, is presented in Figure \ref{fig1}d in the temperature range around $T_{\mathrm{V}}$. The Verwey temperature is inferred for each device as an inflection point in  $R_{\mathrm{DEV}}$(T) dependence; and for the various devices studied here $T_{\mathrm{V}}$ range from 100~K to 110~K. For source and drain electrodes made of the same metal (Cu in this case), $R_{\mathrm{C}}$(source) $\approx$ $R_{\mathrm{C}}$(drain) , besides, $R_{\mathrm{DEV}}$ and $R_{\mathrm{C}}$ have nearly identical temperature dependence (Fig. \ref{fig1}d ).

\begin{figure}
\begin{center}
\includegraphics[clip,width=8cm]{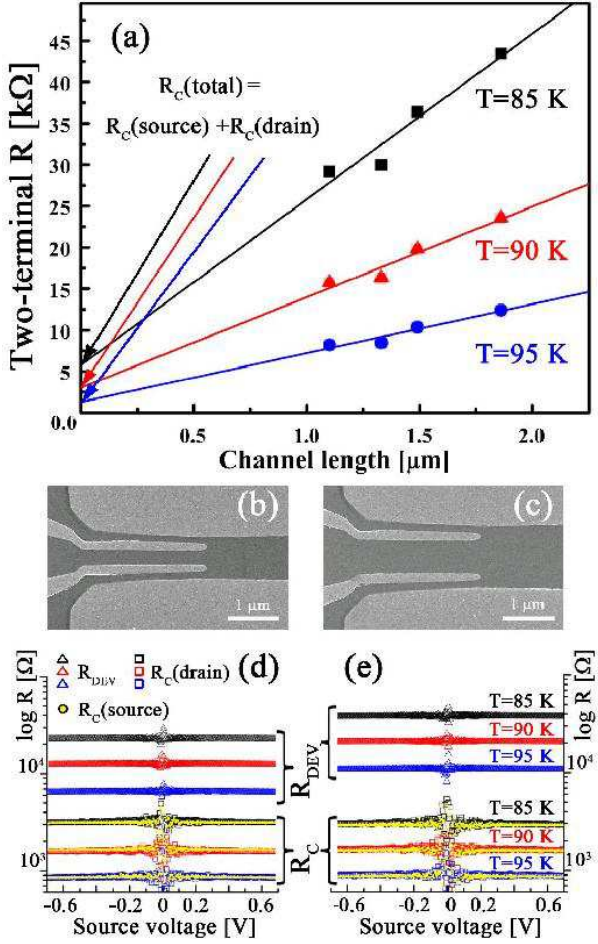}
\end{center}
\vspace{-5mm}
\caption{\small (color) (a) Examples of two-terminal resistance dependence on the channel length at three different temperatures (85~K, 90~K and 95~K) with corresponding linear fits. Extrapolation to zero channel length gives the total contact resistance, R$_C$(total), at each temperature. (b) and (c) show SEM images of devices with different channel lengths. (d) and (e) plot calculated values of device resistances, R$_{DEV}$, and  contact resistances, R$_C$(source) and R$_C$(drain), around zero source voltage for devices in (b) and (c), respectively.}
\label{fig2}
\vspace{-3mm}
\end{figure}

Let us consider the relative contributions of contacts and magnetite channel to the total voltage drop.  Our assumption of linearity in the channel conduction presumes that $R_{\mathrm{DEV}}$ linearly scales with the channel length, $L$, while $R_{\mathrm{C}}$(source) and $R_{\mathrm{C}}$(drain) should remain independent of $L$.   This is supported experimentally.  To demonstrate this, we made a set of devices on the same piece of magnetite film with different channel lengths while all other geometrical parameters (film thickness and the width of source and drain leads) remained exactly the same for all devices. Two representive SEM images of such devices with $L$=1.1~$\mu$m and $L$=1.9~$\mu$m are shown in Fig.~\ref{fig2}b and c, respectively. The total (two-terminal) resistance at each temperature linearly depends on the channel length as demonstrated in Fig.~\ref{fig2}a at several temperatures (85~K, 90~K and 95~K). 

The calculations of contact resistances based on $\ell_{1}$, $\Delta \ell$ and $\ell_2$ for each device show that at each temperature $R_{\mathrm{C}}$(source) and $R_{\mathrm{C}}$(drain) are equal to each other and are the same for devices with different lengths, $L$. It is worth mentioning that $R_{\mathrm{C}}$(source) and $R_{\mathrm{C}}$(drain) do not change upon switching the grounds, {\it i.e.}, exchanging the place of injecting and grounded electrodes. Calculated resistances in Fig. \ref{fig2}d and e represent the data for the devices shown in Fig. \ref{fig2}b and \ref{fig2}c, respectively.  While $R_{\mathrm{C}}$(source) and $R_{\mathrm{C}}$(drain) remain independent of channel length, $R_{\mathrm{DEV}}$ increases as $L$ increases which is obvious from comparison of Fig.~\ref{fig2}d  and Fig.~\ref{fig2}e. All three resistances increase significantly with decreasing the temperature (compare data at 85~K, 90~K and 95~K in Fig.~\ref{fig2}d, e), as will be discussed below in detail. The contact resistance, $R_{\mathrm{C}}$(source) + $R_{\mathrm{C}}$(drain), contributes from 20\% to 13\% of the total two-terminal $R$ for devices with channel lengths ranging from 1~$\mu$m to 2~$\mu$m. 

The increase in total two-terminal resistance with channel length (Fig. \ref{fig2}a) is caused by the increased contribution of $R_{\mathrm{DEV}}$ in longer devices.  Moreover, the extrapolation of a two-terminal $R$ vs $L$ linear fit to zero channel length gives a resistance value very close to the sum of calculated $R_{\mathrm{C}}$(source) and $R_{\mathrm{C}}$(drain) at each temperature.  The latter proves the consistency of our calculations and independence of contact resistances on the channel length within $L$ range investigated in this work.

\begin{figure}
\begin{center}
\includegraphics[clip,width=8cm]{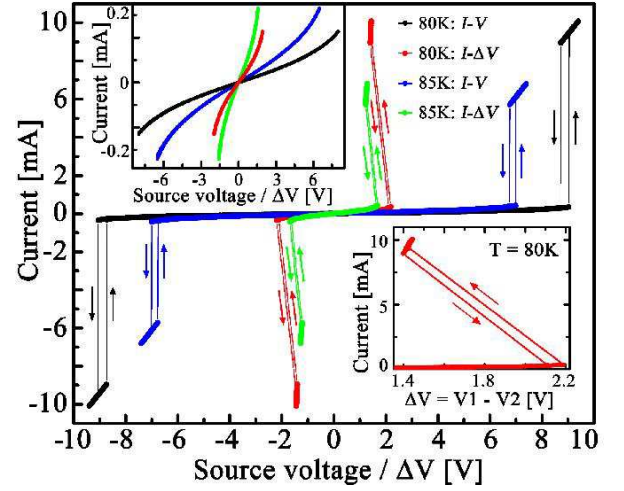}
\end{center}
\vspace{-5mm}
\caption{\small  (color online) Examples of $I$-$V$ and corresponding $I$-$\Delta V$ curves at 80~K and 85~K at the voltage ranges above switching voltage at each temperature. Arrows indicate the direction of the voltage sweep. The pulse sweep parameters are those to demonstrate the small hysteresis in forward and reverse voltage sweeps. Top inset shows $I$-$\Delta V$ curves at the same temperatures, but at the voltage range below switching voltage.
Bottom inset zooms in to $I$-$\Delta V$ curve at 80~K around transition point to demonstrate the discontinuity in measured $\Delta V$ value.}
\label{fig3}
\vspace{-3mm}
\end{figure}

At temperatures below $T_{\mathrm{V}} \sim 105K$,  the current-voltage characteristics, $I$-$V$, show Ohmic behavior at low source voltage range ($<$ 1V), while start to exhibit nonlinearities at higher voltages, symmetrical for positive and negative source voltages.  Examples at two selected temperatures (80~K and 85~K) are shown in the top inset of Figure~\ref{fig3}. Upon further increasing source voltage $I$-$V$ curves show a sharp jump in current (Fig.~\ref{fig3}) as soon as the source voltage reaches a critical switching value, $V_{SW}$, at a certain temperature as described in detail in \cite{Our_magnetite_2008, 2009_PRB_hyster}.  This is a transition from high resistance (Off) state to a state with much lower resistance (On) state.  Note the two-order of magnitude difference in current after transition by comparing $I$-$V$ curves before (top inset) and after transition in Fig.~\ref{fig3}.

Corresponding $I$ vs $\Delta V=V_{1}-V{_2}$ plots (Fig. \ref{fig3}) have much lower switching $\Delta V_{SW}$ values and reveal at a transition point not only a discontinuity in current, but also in the measured $\Delta V$ value, which decreases in absolute magnitude (Fig. \ref{fig3} bottom inset).  Since $\Delta V=V_{1}-V_{2}$, in general, reflects properties of magnetite channel without contact effects, the discontinuity (jump) in $\Delta V$ at a transition point can be explained as a sudden decrease in device resistance, $R_{\mathrm{DEV}}$.

\begin{figure}
\begin{center}
\includegraphics[clip,width=8.5cm]{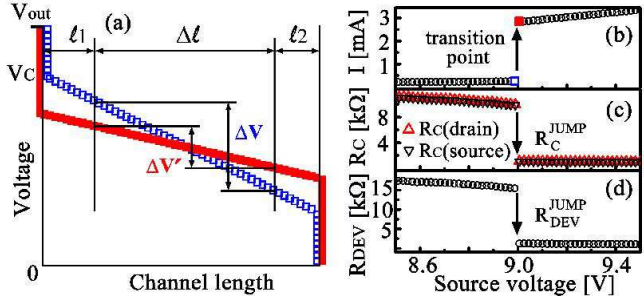}
\end{center}
\vspace{-5mm}
\caption{\small  (color online) (a) Schematic diagram of voltage distribution along the channel at a transition point (blue open squares) and right after transition (red closed squares). V$_{\mathrm{C}}$ denotes the voltage drop at the interfaces. (b) A fragment of $I$-$V$ curve at 85~K in the vicinity of the transition. Blue open square marks the transition point and red closed square shows a point right after transition; blue and red squares in (a) corresond to the voltage distribution over the channel at these points.  (c) and (d) are the voltage dependences of $R_{\mathrm{C}}$(source), $R_{\mathrm{C}}$(drain) and $R_{\mathrm{DEV}}$, respectively, demonstrating the abrupt decreases (jumps) in all three resistances at the transition point.}
\label{fig4scheme}
\vspace{-3mm}
\end{figure}

Now let us turn to the quantitative description of contact effects at the onset of the field-induced transition. Calculations of $R_{\mathrm{C}}$(source), $R_{\mathrm{DEV}}$ and $R_{\mathrm{C}}$(drain) show that at a transition point the voltage drops at the contacts, $V_{\mathrm{C}}$(source) and $V_{\mathrm{C}}$(drain), increase in absolute value, while $\Delta$V decreases. Compare the blue open squares (at the transition point) and red closed squares (at the next point after transition) in Fig.~\ref{fig4scheme}a, which depict voltage distribution over the channel length. From this sketch the decrease in $\Delta$V value is also clearly visible.

Although $V_{\mathrm{C}}$(source) and $V_{\mathrm{C}}$(drain) increase, due to the overall increased current, $R_{\mathrm{C}}$(source) and $R_{\mathrm{C}}$(drain) actually {\it decrease} upon passing through the transition point.  Fig.~\ref{fig4scheme}b,c, and d explicitly demonstrates these decreases (jumps) in $R_{\mathrm{DEV}}$ and $R_{\mathrm{C}}$ at a transition point, denoted further as $R^{jump}_{\mathrm{DEV}}$ and $R^{jump}_{\mathrm{C}}$. For source and drain electrodes made of Cu, $R_{\mathrm{C}}$(source) and $R_{\mathrm{C}}$(drain) jumps at the transition point are equal to each other and remain unchanged in the experiments on exchanging the grounds.

Note that at the transition point {\it both} device and source and drain contact resistances decrease abruptly.  This behavior is distinct from the one for other systems exhibiting voltage-driven transitions  (such as manganites and doped SrTiO$_{3}$). For these systems the leading role of oxygen vacancies drift under applied bias was demonstrated\cite{2006_Nature_SrTiO3_Waser, 2007_oxygen_diff_Ignatiev}, and source and drain contact resistances show variations of opposite sign \cite{2007_PRL_mechanism}; {\it i.e.}, while source contact resistance increases, the drain contact resistance decreases.  Since this is inconsistent with our observations, magnetite clearly must exhibit a different switching mechanism.

\begin{figure}
\begin{center}
\includegraphics[clip,width=8cm]{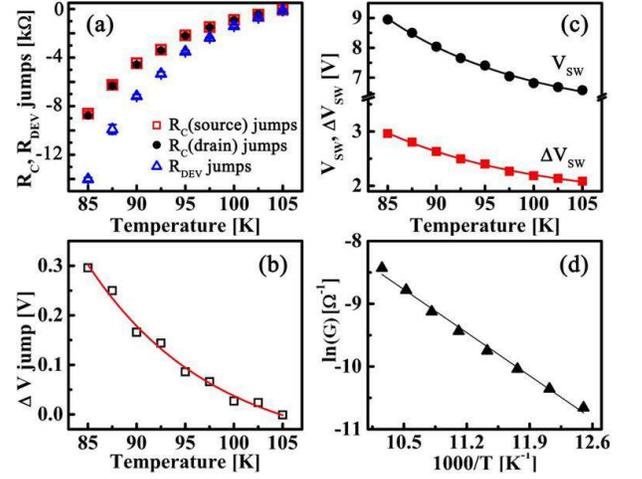}
\end{center}
\vspace{-5mm}
\caption{\small (color online)   Temperature dependences of  (a) the jumps in  $R_{\mathrm{C}}$(source), $R_{\mathrm{C}}$(drain) and in $R_{\mathrm{DEV}}$ at a transition point; each point represents an average over 8 independent measurements (left/right voltage pairs, different grounds, positive/negative switching voltages), standard deviation is within the symbol size (b) jumps in $\Delta$V (squares) at a transition point and its exponential fit (solid line) (c) $V_{SW}$ and $\Delta V_{SW}$ with corresponding exponential fits and (d) conductance ($1/R_{\mathrm{DEV}}$) of magnetite channel in Arrhenius coordinates and its linear fit.}
\label{fig5}
\vspace{-3mm}
\end{figure}

As temperature decreases, the jumps in contact resistance, $R^{jump}_{\mathrm{C}}$, and device resistance, $R^{jump}_{\mathrm{DEV}}$, remain negative, but increase in absolute magnitude (Fig.~\ref{fig5}a). Upon approaching the temperature when switching is not observed (T $\sim$ $T_{\mathrm{V}}$ \cite{Our_magnetite_2008, 2009_PRB_hyster}), $R^{jump}_{C}$ and $R^{jump}_{DEV}$ approach zero. The jump in $\Delta V$ is also temperature dependent and its magnitude exponentially decays with the temperature, approaching zero at $T_{\mathrm{V}}$. $V_{SW}$ and $\Delta V_{SW}$  depend on temperature exponentially as well, as demonstrated in Fig.~\ref{fig5}c.


To explain the temperature dependence of the parameters related to the observed transition, {\it i.e.}, jumps in current, $\Delta V$, $R_{\mathrm{C}}$ and $R_{\mathrm{DEV}}$ (Fig.~\ref{fig5}), we should review the properties of magnetite below Verwey temperature, since the transition is only observed  below $T_{\mathrm{V}}$. We will consider the Verwey transition physics in magnetite from the electronic structure point of view.  While magnetite has strong electron correlations, it is believed that a band-type description of its electronic structure is a reasonable approximation \cite{2007_spin_PES_review}, with transport being dominated by low-lying electronic states near an effective Fermi energy, $E_{\mathrm{F}}$.

The electronic structure of magnetite has been probed extensively by photoelectron and scanning tunneling spectroscopies and band-structure was calculated using different methods \cite{1995_PES_gap_closed, 1984_highT_band_struc_calc, 2006_PRB_STM_film_surface}.  Above $T_{\mathrm{V}}$ there is a finite (non-zero) density of states (DOS) around $E_{\mathrm{F}}$, which is dominated by Fe 3$d$ states of the $B$-site sublattice of cubic structure.  Below $T_{\mathrm{V}}$ the DOS near $E_{\mathrm{F}}$ exhibits a clear gap, causing two orders of magnitude increase in resistivity at $T_{\mathrm{V}}$ \cite{1995_PES_gap_closed, 2005_PES_EuroPhys}.

Recently, it has been theoretically predicted that a gap to charged excitations (charge gap) in correlated insulators can be closed by applying external electric field, resulting in field-induced metal-insulator transition \cite{2003_PRL_Mott_break, 2008_PRB_theory}. Several systems exhibit this behavior, for example, charge-ordered state of complex manganese oxides \cite{1997_Xray_MIT_Tokura,1997_Tokura_first}, as well as 1D cuprates \cite{2001_PRB_1D_Mott_break} and 2D nickelates \cite{1999_PRL_2D_Mott_break}. 

Our data on magnetite seem to be another experimental observation
consistent with this sort of gap closure by electric field.  The
absence of hysteresis in forward and reverse bias sweeps, meaning that
metallic state persists only if applied voltage (electric field)
exceeds a critical value, $|V|>|V_{SW}|$ , is expected from this mechanism 
and is indeed observed in Fe$_{3}$O$_{4}$ (see Ref.~\cite{2009_PRB_hyster}
for details).  The jump in $|R_{\mathrm{DEV}}|$ at the transition point is a natural 
consequence of the gap closure.  The increase in the absolute value of
$R^{\mathrm{jump}}_{\mathrm{DEV}}$ as $T$ decreases (Fig.~\ref{fig5}a)
is also easily explained, as at lower temperatures there is a
transition from more insulating state to the same metallic state with
zero-size gap. Accompanying drops of contact resistances, $R^{jump}_{C}$, are 
direct consequences of gap closure and, thus, the change in the position of magnetite $E_{\mathrm{F}}$ relative to the $E_{\mathrm{F}}$ of the contact metal.

The exponential dependences of the above parameters, particularly that shown in Fig.~\ref{fig5}d, imply thermally activated transport below $T_{\mathrm{V}}$.  Indeed, plotting the inverse of $R_{\mathrm{DEV}}$ (in the Ohmic regime near zero bias) in Arrhenius coordinates gives a straight line in a given $T$ range (Fig.\ref{fig5}d). The activation energy, $E_{\mathrm{a}}$, inferred from these data on the magnetite channel (device) lies in 85-89 meV range for several devices.  These values  match well with the size of the gap below $T_{\mathrm{V}}$, inferred from photoemission  \cite{1995_PES_gap_closed, 2005_PES_EuroPhys} and optical \cite{1998_optical_gap_Tokura, 2005_THz_cond} spectroscopies data.  This suggests that transport below $T_{\mathrm{V}}$ involves charge carriers thermally activated over the gap.

In conclusion, by doing four-terminal experiments at magnetite thin films below $T_{\mathrm{V}}$ we quantitatively separate the contributions of each electrode and the magnetite channel before and after the electric field driven transition.  For devices of increasing channel lengths we demonstrate the increase in total resistance to be caused by increased contribution of the magnetite channel, while contact resistances are unchanged for all channel lengths within 1 to 2 $\mu$m range. At all temperatures the transition is observed ($T< T_{\mathrm{V}}$), contact resistances of {\it both} source and drain electrodes and the resistance of magnetite channel decrease abruptly at the transition point. This behavior is consistent with the mechanism of charge gap closure by electric field predicted in theory \cite{2003_PRL_Mott_break, 2008_PRB_theory}. To further explore the field-driven switching mechanism in magnetite, the effect of contact metals with different work functions is currently under study. In the framework of the charge gap closure mechanism,  the magnitude of contact resistance jumps at a transition point, $R^{jump}_{C}$, are expected to be dependent on the work function of the contact metal according to the relative alignment of metal Fermi level and effective Fermi level of magnetite.

This work was supported by the US Department of Energy grant
DE-FG02-06ER46337.  DN also acknowledges the David and Lucille Packard
Foundation and the Research Corporation. RGSS and
IVS acknowledge the Science Foundation of Ireland grant 06/IN.1/I91.


\begin{thebibliography}{10}

\bibitem{1998_Tokura_MIT_review}
M.~Imada, A.~Fujimori, and Y.~Tokura,
\newblock Rev. Mod. Phys. {\bf 70}, 1039 (1998).

\bibitem{1939_Verwey_first_Nature}
E.~J.~W.~Verwey,
\newblock Nature {\bf 144}, 327 (1939).

\bibitem{2002_Walz_review}
F.~Walz,
\newblock J. Phys.: Condens. Matter. {\bf 14}, R285 (2002).

\bibitem{1954_mag_el_lowT}
B.~A.~Calhoun,
\newblock Phys. Rev. {\bf 94}, 1577.

\bibitem{2007_planar_Hall}
Y.~Bason, L.~Klein, H.~Q.~Wang, J.~Hoffman, X.~Hong, V.~E.~Henrich, and C.~H.~Ahn,
\newblock J. Appl. Phys. {\bf 101}, 09J507 (2007).

\bibitem{1982_el_trans_MO}
J.~M.~Honig,
\newblock J. Solid State Chem. {\bf 45}, 1 (1982).

\bibitem{2007_spin_PES_review}
M.~Fonin, Yu.~S.~Dedkov, R.~Pentcheva, U.~R\"{u}diger, and G.~G\"{u}ntherodt,
\newblock J. Phys.: Condens. Matter {\bf 19}, 315217 (2007).

\bibitem{1987_polaron_cond}
L.~Degiorgi, P.~Wachter, and D.~Ihle,
\newblock Phys. Rev. B {\bf 35}, 9259.

\bibitem{2004_CO_OO}
I.~Leonov, A.~N.~Yaresko, V.~N.~Antonov, M.~A.~Korotin, and V.~I.~Anisimov,
\newblock Phys. Rev. Lett. {\bf 93}, 146404 (2004).

\bibitem{1984_highT_band_struc_calc}
A.~Yanase and K.~Siratori,
\newblock J. Phys. Soc. Jpn. {\bf 53}, 312 (1984).

\bibitem{Our_magnetite_2008}
S.~Lee, A.~Fursina, J.~T.~Mayo, C.~T.~Yavuz, V.~L.~Colvin, R.~G.~S.~Sofin, I.~V.~Shvets, and D.~Natelson,
\newblock Nature Mater. {\bf 7}, 130 (2008).

\bibitem{2008_APL_HAR_Cr}
A.~Fursina, S.~Lee, R.~G.~S.~Sofin, I.~V.~Shvets, and D.~Natelson,
\newblock Appl. Phys. Lett. {\bf 92}, 113102 (2008).

\bibitem{2009_PRB_hyster}
A.~A.~Fursina, R.~G.~S.~Sofin, I.~V.~Shvets, and D.~Natelson,
\newblock Phys. Rev. B {\bf 79}, 245131 (2009).

\bibitem{1969_Tinduced_MIT_Fe3O4_1}
P.~J. Freud and A.~Z. Hed,
\newblock Phys. Rev. Lett. {\bf 23}, 1440 (1969).

\bibitem{1969_Tinduced_MIT_Fe3O4_2}
T.~Burch, P.~P.~Craig, C.~Hedrick, T.~A.~Kitchens, J.~I.~Budnick, J.~A.~Cannon, M.~Lipsicas, and D.~Mattis,
\newblock Phys. Rev. Lett. {\bf 23}, 1444 (1969).


\bibitem{Shvets_backscat}
A.~Koblischka-Veneva, M.~R.~Koblischka, Y.~Zhou, S.~Murphy, F.~M\"{u}cklich, U.~Hartmann, I.~V.~Shvets,
\newblock J. Magn. Magn. Mater. {\bf 316}, e663 (2007).

\bibitem{Shvets_high_res_Xray}
S.~K. Arora, R.~G.~S. Sofin, I.~V. Shvets, and M.~Luysberg,
\newblock J. Appl. Phys. {\bf 100}, 073908 (2006).

\bibitem{2006_Nature_SrTiO3_Waser}
K.~Szot, W.~Speier, G.~Bihlmayer, and R.~Waser,
\newblock Nat. Mater. {\bf 5}, 312 (2006).

\bibitem{2007_oxygen_diff_Ignatiev}
Y.~B. Nian, J.~Strozier, N.~J. Wu, X.~Chen, and A.~Ignatiev,
\newblock Phys. Rev. Lett. {\bf 98}, 146403 (2007).

\bibitem{2007_PRL_mechanism}
M.~Quintero, P.~Levy, A.~G. Leyva, and M.~J.~Rozenberg,
\newblock Phys. Rev. Lett. {\bf 98}, 116601 (2007).

\bibitem{1995_PES_gap_closed}
A.~Chainani, T.~Yokoya, T.~Morimoto, T.~Takahashi, and S.~Todo,
\newblock Phys. Rev. B {\bf 51}, 17976 (1995).

\bibitem{2006_PRB_STM_film_surface}
K.~Jordan, A.~Cazacu, G.~Manai, S.~F.~Ceballos, S.~Murphy, and I.~V.~Shvets,
\newblock Phys. Rev. B {\bf 74}, 085416 (2006).

\bibitem{2005_PES_EuroPhys}
D.~Schrupp, M.~Sing, M.~Tsunekawa, H.~Fujiwara, S.~Kasai, A.~Sekiyama, S.~Suga, T.~Muro, V.~A.~M.~Brabers, and R. Claessen,
\newblock Europhys. Lett. {\bf 70}, 789 (2005).

\bibitem{2003_PRL_Mott_break}
T.~Oka, R.~Arita, and H.~Aoki,
\newblock Phys. Rev. Lett. {\bf 91}, 066406 (2003).

\bibitem{2008_PRB_theory}
N.~Sugimoto, S.~Onoda, and N.~Nagaosa,
\newblock Phys. Rev. B {\bf 78}, 155104 (2008).

\bibitem{1997_Xray_MIT_Tokura}
V.~Kiryukhin, D.~Casa, J.~P.~Hill, B.~Keimer, A.~Vigliante, Y.~Tomioka and Y.~Tokura,
\newblock Nature {\bf 386}, 813 (1997).

\bibitem{1997_Tokura_first}
A.~Asamitsu, Y.~Tomioka, H.~Kuwahara, and Y.~Tokura,
\newblock Nature {\bf 388}, 50 (1997).

\bibitem{2001_PRB_1D_Mott_break}
Y.~Taguchi, T.~Matsumoto, and Y.~Tokura,
\newblock Phys. Rev. B {\bf 62}, 7015 (2000).

\bibitem{1999_PRL_2D_Mott_break}
S.~Yamanouchi, Y.~Taguchi, and Y.~Tokura,
\newblock Phys. Rev. Lett. {\bf 83}, 5555 (1999).

\bibitem{1998_optical_gap_Tokura}
S.~K.~Park, T.~Ishikawa, and Y.~Tokura,
\newblock Phys. Rev. B {\bf 58}, 3717 (1998).

\bibitem{2005_THz_cond}
A.~Pimenov, S.~Tachos, T.~Rudolf, A.~Loidl, D.~Schrupp, M.~Sing, R.~Claessen, and V.~A.~M Brabers,
\newblock Phys. Rev. B {\bf 72}, 035131 (2005).

\end{thebibliography}
\end{document}